\documentclass[conference]{IEEEtran}
\IEEEoverridecommandlockouts

\usepackage{cite}
\usepackage{amsmath,amssymb,amsfonts}
\usepackage{algorithmic}
\usepackage{graphicx}
\usepackage[T1]{fontenc}
\usepackage{xcolor}
\def\BibTeX{{\rm B\kern-.05em{\sc i\kern-.025em b}\kern-.08em
    T\kern-.1667em\lower.7ex\hbox{E}\kern-.125emX}}

\usepackage{fontawesome5}
\usepackage{etoolbox}
\usepackage[percent]{overpic}
\usepackage{tabularx}
\usepackage{booktabs}
\usepackage{stfloats}
\usepackage[hidelinks,pdfencoding=auto]{hyperref}
\begin{document}

\title{A 2.4 GHz LC-VCO Fractional-N Phase Locked Loop Open-Source Design in 130-nm BiCMOS
\thanks{
This is the author’s version of the article that has been accepted for presentation at the International Conference on Synthesis, Modeling, Analysis and Simulation Methods, and Applications to Circuit Design (SMACD) 2026.
}
\thanks{\textcopyright~2026 IEEE. Personal use of this material is permitted. Permission from IEEE must be obtained for all other uses, 
in any current or future media, including reprinting/republishing this material for advertising or promotional purposes, creating new collective works, for resale or redistribution to servers or lists, or reuse of any copyrighted component of this work in other works.
}}
 

\author{
\IEEEauthorblockN{Manimohan Thiriloganathan\textsuperscript{1},
Shenal Ranasinghe\textsuperscript{1},
Avishka Herath\textsuperscript{1},
Rajinthan Rameshkumar\textsuperscript{1},\\
Hansa Marasinghe\textsuperscript{1},
Anjana Viduranga\textsuperscript{1},
Gayangana Leelarathne\textsuperscript{2} and
Kithmin Wickremasinghe\textsuperscript{3}} \\
\IEEEauthorblockA{$^{1}$Department of Electronic and Telecommunication Engineering, University of Moratuwa, Sri Lanka}
\IEEEauthorblockA{$^{2}$School of Electrical Engineering, Aalto University, Finland}
\IEEEauthorblockA{$^{3}$Department of Electrical and Computer Engineering, University of British Columbia, Canada}
}

\maketitle

\begin{abstract}
Radio frequency (RF) integrated circuit design using the open-source complementary Metal-Oxide semiconductor (CMOS) ecosystem, such as for phase-locked loops (PLLs), is limited by the absence of reliable passive device models, particularly on-chip spiral inductors. Consequently, prior work relies on ring-oscillator-based voltage-controlled oscillators (VCOs) with degraded phase noise performance. This work presents a 2.4~GHz type-II fractional-N PLL implemented in the IHP SG13G2 130 nm BiCMOS open-source technology. The proposed design employs a cross-coupled differential LC-VCO integrated with a custom-designed spiral inductor, developed using an open-source electromagnetic modelling workflow in OpenEMS. The optimized inductor achieves 4~nH inductance with a quality factor of 16.8 at 2.45~GHz. The LC-VCO sensitivity is approximately 120~MHz/V while the PLL phase noise is -100.8~dBc/Hz at 1~MHz offset. The complete PLL is realized using a fully open-source electronic design automation (EDA) flow, occupying a total area of 930~$\mu\text{m}~\times~$666$~\mu\text{m}$ ($\approx$ 0.619$~\text{mm}^2$) and consuming 12.73$~\text{mW}$, demonstrating the feasibility of RF integrated circuit design in an open-source CMOS IC design ecosystem. 
\end{abstract}


\section{Introduction}

Phase-locked loops are indispensable components in modern wireless communication systems, serving as the core of frequency synthesizers, clock generators, and timing recovery circuits \cite{11383160}. With the proliferation of Internet of Things (IoT) devices, there is a significant push towards low-power, high-performance PLL architectures that can be implemented using open-source methodologies to reduce design costs and increase accessibility.

Despite the critical role of these circuits, the open-source chip design environment faces technical hurdles. As noted in recent research \cite{razavi_edu, 11383160}, the absence of spiral inductor models in open-source process design kits (PDKs) hampers traditional LC-based designs, promoting inductor-less ring oscillators. Furthermore, available open-source EDA tools currently do not support direct phase noise simulation, necessitating manual transient analysis and processing to estimate performance\cite{11383160}. To address these challenges, there is an open need for work adopting a fully open-source RF design methodology utilizing open-source PDKs. Such workflows would demonstrate pathways for students and researchers to overcome the barriers of traditional licensing and proprietary tools.
\begin{figure}[t!]
    \centering
    \includegraphics[width=1\linewidth]{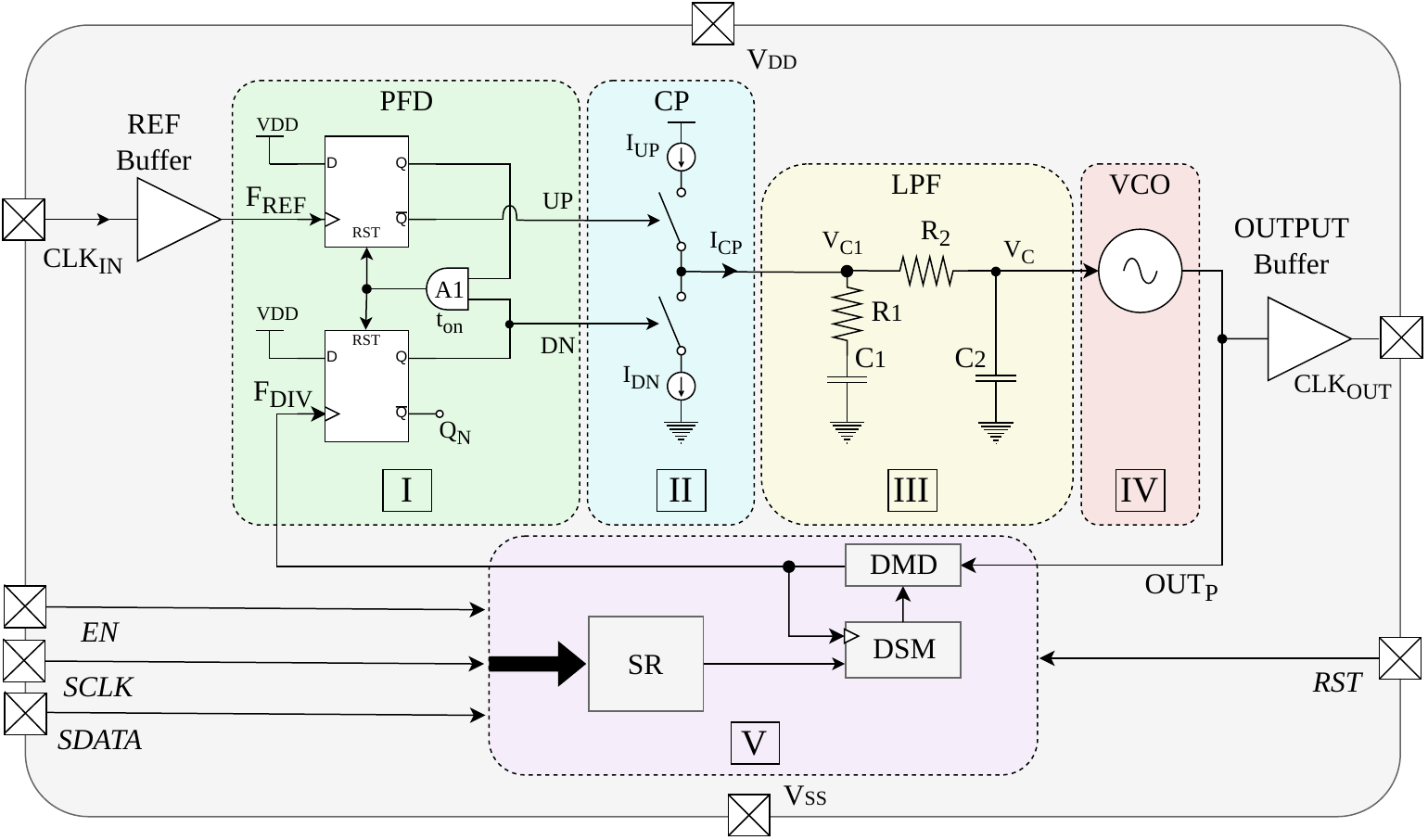}
    \caption{The LC-VCO-based type-II fractional-N PLL design architecture.} 
    \label{fig:placeholder}
    \vspace{-0.6cm}
\end{figure}

This paper presents the design of a 2.4~GHz type-II fractional-N PLL leveraging the IHP-Leibniz Institute's SG13G2 130 nm SiGe BiCMOS open-source PDK \cite{ihp_pdk2}. The design is specifically aimed for the 2.4 GHz industrial, scientific, and medical (ISM) band, making it suitable for Bluetooth low energy (BLE) applications \cite{ref6}. BLE systems require stringent phase noise performance and fast settling times (typically $<$150~$\mu$s) to maintain stable frequency hopping and high-quality data transmission \cite{ble_standard}.

A key contribution of this work is the adoption of an entirely open-source EDA flow. The design utilizes \textit{Xschem} for schematic capture, \textit{ngspice} for transient and noise simulations, and \textit{KLayout} for layout generation. The core of the system is a cross-coupled differential LC-VCO, where a custom-designed 4~nH inductor was modelled and integrated. Unlike traditional closed flows, this design employs open-source inductor modelling strategies, utilizing electromagnetic (EM) solvers like \textit{OpenEMS} to characterize the device within the SG13G2 backend-of-line (BEOL) metal stack. This approach demonstrates that RF blocks, previously reserved for proprietary tools, can now be developed within the open-source CMOS IC design ecosystem. To the best of our knowledge, this is the first-ever work that demonstrates a fully open-source CMOS PLL design with an integrated on-chip spiral inductor. Further details are available in the GitHub repository: \href{https://github.com/Manimohan05/SG13G2_2.4GHz_LC_VCO_FPLL}{\small https://github.com/Manimohan05/SG13G2\_2.4GHz\_LC\_VCO\_FPLL}.

\section{Inductor Design and Simulation Workflow}

\begin{figure}[b!]
    \vspace{-0.2cm}
    \centering
    \includegraphics[width=0.9\columnwidth]{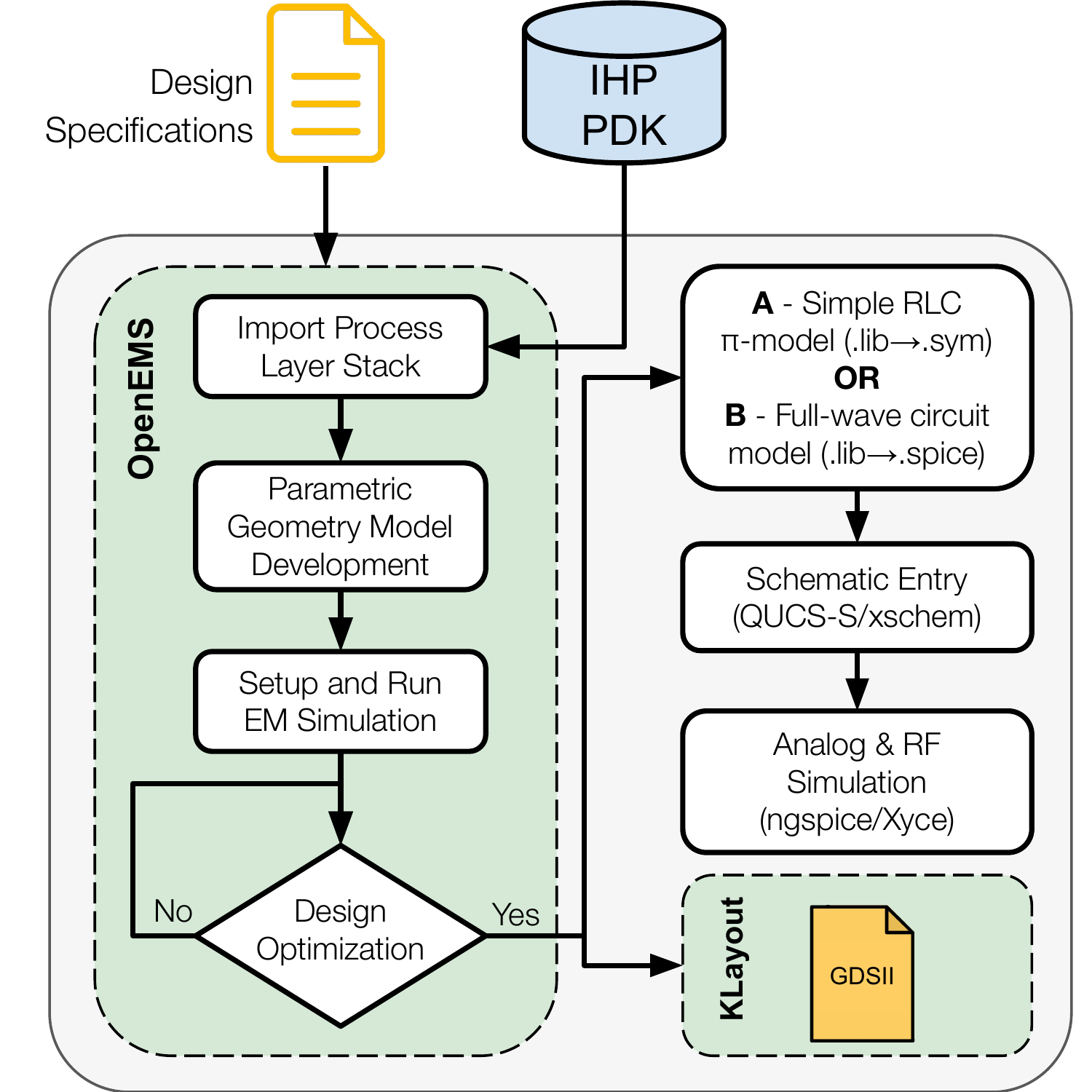}
    \caption{\textbf{Workflow for parametric RF inductor design.} This represents the main steps involved in the design optimization workflow of a spiral inductor using open-source tools such as \textit{OpenEMS}, \textit{ngspice} and \textit{Xschem}.}
    \label{fig:workflow}
\end{figure}

The VCO constitutes the core building block of the
proposed PLL, and its phase-noise performance is fundamentally governed by the quality factor ($Q_{\text{diff}}$) of the LC tank. Because the inductor is the dominant loss element in the tank, maximizing its $Q_{\text{diff}}$ at the oscillation frequency is a first-order design objective. Integrated inductors in a CMOS process are realized as metal spirals whose critical geometric parameters, namely the trace width ($W$), turn spacing ($S$), number of turns ($N$), and outer radius ($R_{\text{out}}$), present well-known trade-offs~\cite{razavi2012}. A key contribution of this work was demonstrating an open-source spiral inductor design and optimization workflow, thus enabling reproducible RF component development without reliance on proprietary electromagnetic solvers. The complete design and optimization workflow that was developed is summarized in Fig.~\ref{fig:workflow}.

\begin{figure}[!b]
\vspace{-0.5cm}
\centering
\includegraphics[width=0.95\columnwidth]{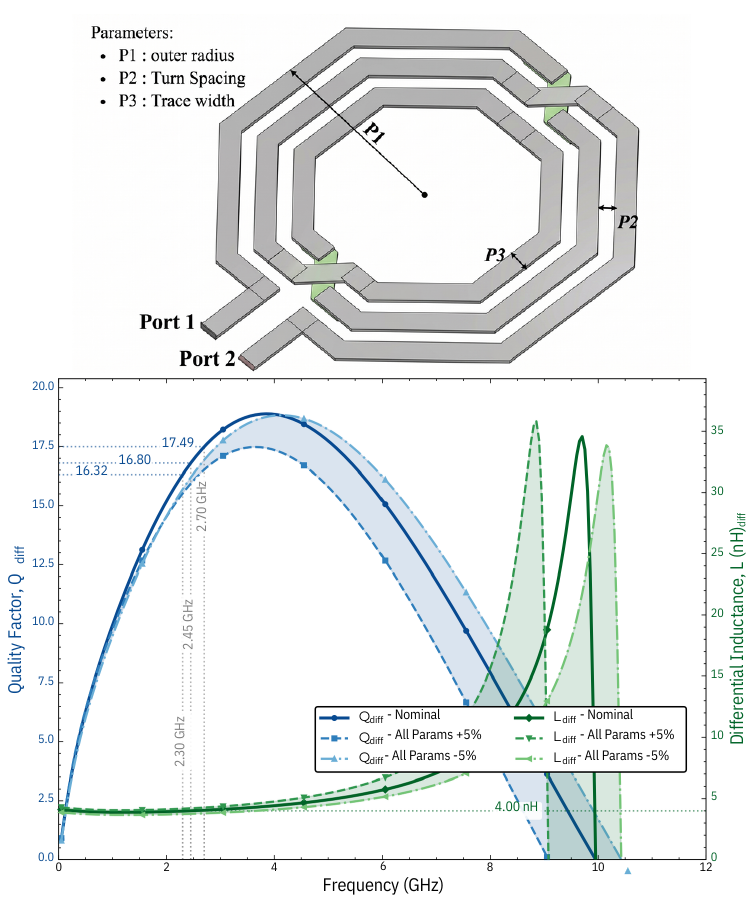}
\vspace{-0.2cm}
\caption{\textbf{Results of inductor design:} \textbf{(top)} \textbf{optimized parametric 3D layout:} $R_{\text{out}}~(P_{1}) = 218~\mu$m, $S~(P_{2}) = 14~\mu$m, $W~(P_{3}) = 30~\mu$m; \textbf{(bottom)} \textbf{parametric study:} plot of $Q_{\text{diff}}$ and $L_{\text{diff}}$ vs frequency. The Q-factor curve (solid, dark blue) rises monotonically to a peak of approximately 18.9 near 3.8~GHz before declining as substrate and skin effect losses dominate, while the differential inductance (solid, dark green) remains nearly flat across the 2--4~GHz band and rolls off sharply approaching the SRF near 10~GHz. Further, sensitivity analysis is demonstrated by the $+$5\% perturbation (dashed, medium blue for $Q_{\text{diff}}$; dashed, medium green for $L_{\text{diff}}$) and the $-$5\% perturbation (dash-dot, light blue for $Q_{\text{diff}}$; dash-dot, light green for $L_{\text{diff}}$).}
\label{fig:inductor_parametric}
\end{figure}

For our design, a three-turn ($N=3$) symmetric octagonal spiral geometry was adopted, offering additional layout degrees of freedom as opposed to a square or circular alternative. The inductor is modelled using a parametric design that interfaces directly with the IHP SG13G2 BEOL metal stack. The main spiral is placed on \textit{TopMetal\,2}, the thickest available metal layer chosen to minimize series resistance, while underpasses are routed on \textit{TopMetal\,1} with \textit{TopVia2} providing the interlayer transitions. In the parametric design shown in Fig.~\ref{fig:inductor_parametric} (top), three independent design variables were used to define the inductor geometry: outer radius $P_{1} = R_{\text{out}}$, turn spacing $P_{2} = S$, and trace
width $P_{3} = W$. Initial estimates for these parameters were derived from modified Wheeler's formula~\cite{Mohan1999} to bound the simulation design space.

The parametric model is exported to \textit{OpenEMS}, an open-source finite-difference time-domain (FDTD) electromagnetic solver~\cite{openems}. A Gaussian-pulse excitation covering 0--12~GHz is applied at the differential ports, and the complete substrate stack, including silicon bulk, oxide inter-metal dielectrics, and passivation, is imported from the PDK layer definitions. To accurately model conductor losses, the simulation utilizes a \textit{refined\_cellsize} of 0.5 $\mu$m. Automated parameter sweeps vary $P_{1}$, $P_{2}$, and $P_{3}$ to maximize $Q_{\text{diff}}$ at 2.45\,GHz while constraining $L_{\text{diff}}$ to the 4~nH target within 2\% tolerance. Raw two-port $S$-parameters are de-embedded via a two-step open-short procedure~\cite{5235929} to remove pad and feed line parasitics.
The resulting de-embedded $Z$-parameters yield the differential impedance $Z_{\text{diff}} = z_{11} - z_{12} - z_{21} + z_{22}$, from which the intrinsic inductance, $L_{\text{diff}}$, and quality factor, $Q_{\text{diff}}$, are extracted using: \vspace{-0.1cm}
\begin{equation*}
    L_{\text{diff}} = \frac{\text{Im}(Z_{\text{diff}})}{\omega}, \qquad
    Q_{\text{diff}} = \frac{\text{Im}(Z_{\text{diff}})}{\text{Re}(Z_{\text{diff}})}.
\end{equation*}

The optimized geometry shown in  Fig.~\ref{fig:inductor_parametric} yields $L_{\text{diff}} = 4.000$~nH and
$Q_{\text{diff}} = 16.80$ at 2.45~GHz. Further, a sensitivity analysis of $P_{1}$, $P_{2}$, $P_{3}$ is overlaid with shaded regions, with $Q_{\text{diff}}$ varying between 16.32 and 17.49 across across the 2.3--2.7~GHz band and $L_{\text{diff}}$ remaining within 3.9--4.1~nH across the 2.3--2.7~GHz band, confirming robust operation of the design. The self-resonance frequency (SRF) lies well above the operating frequency, ensuring stable integration as well. 

\begin{figure*}[t!]
    \centering
    \includegraphics[width=\textwidth]{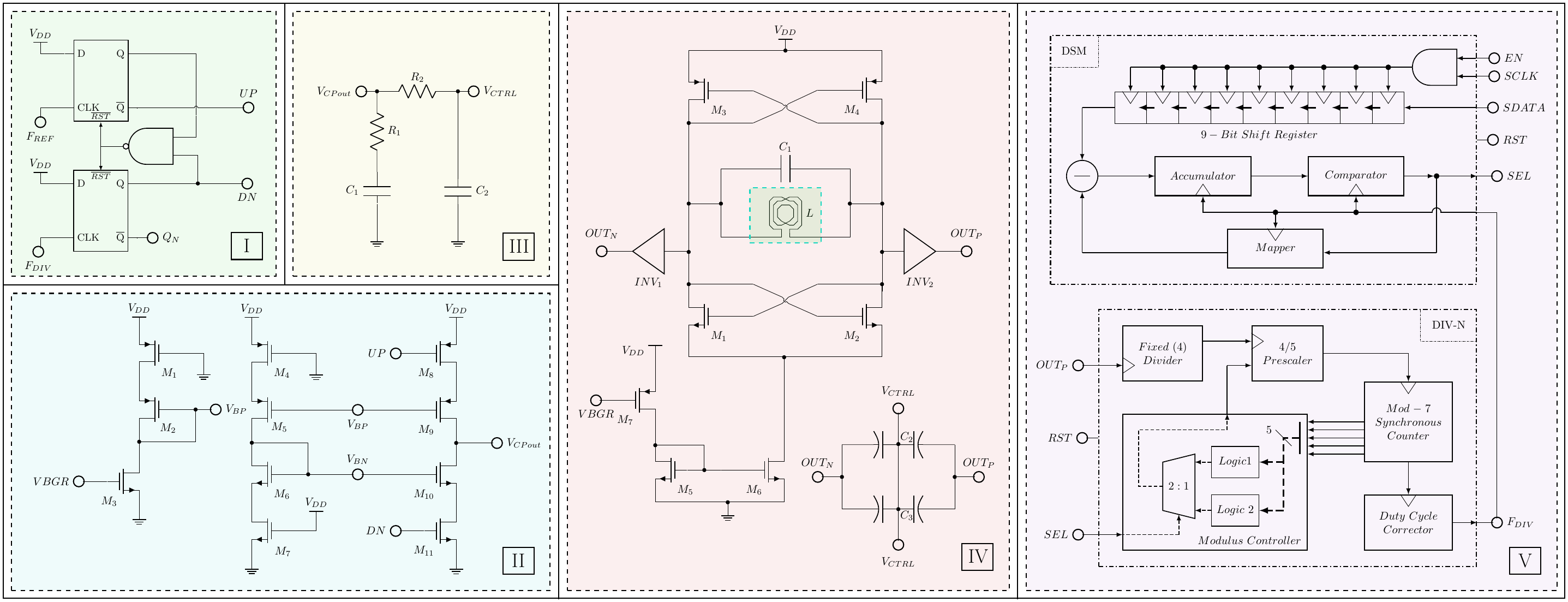}
    
    \caption{\textbf{Block diagram of the proposed fractional-N PLL architecture.} The system consists of five main blocks: (I) phase–frequency detector (PFD), (II) charge pump (CP), (III) loop filter (LF), (IV) LC cross-coupled VCO, and (V) programmable fractional-N divider based on a delta sigma ($\Delta \Sigma$) modulator.} 
    \label{fig:pll_blocks}
    \vspace{-0.5cm}
\end{figure*}

\section{Fractional-N PLL Design Methodology}

\subsection{Phase Frequency Detector (PFD)}
The PFD is responsible for detecting the phase and frequency deviation between the reference clock ($F_{REF}$) and the divided feedback clock ($F_{DIV}$). The design in Fig.~\ref{fig:pll_blocks} utilizes two edge-triggered D-flip-flops (DFFs). NAND gate in the feedback path monitors the UP and DN outputs to reset both DFFs, which prevents the PFD from locking up and provides a tunable delay to mitigate the dead-zone \cite{razavi2012}.

\subsection{Charge Pump (CP) and Loop Filter (LF)}
The CP, illustrated in Fig.~\ref{fig:pll_blocks}, translates UP and DN pulses from the PFD into discrete current pulses (sourcing or sinking) to adjust the control voltage ($V_{CTRL}$). The subsequent LF integrates these charge packets into a continuous control voltage for the VCO, while also suppressing high-frequency reference spurs generated by the PFD switching activity. 

\subsection{LC Voltage Controlled Oscillator (LC-VCO)}

A type-IV cross-coupled topology \cite{ham2001,razavi2012} is used to achieve low phase noise, a tuning range above 3.3\% (2.4--2.48~GHz), and stable startup across the control voltage range. A cross-coupled NMOS pair ($M_1$, $M_2$) provides the negative resistance to cancel tank losses, while a PMOS pair ($M_3$, $M_4$) improves transconductance and helps startup under low supply voltage. 
Two parallel varicaps are used to provide a capacitance change of 70--200~fF, which was obtained with a voltage change of 1.2~V in $V_{CTRL}$.
The transistor sizes ($M_1$–$M_4$) are chosen to ensure enough transconductance for reliable startup near 0.7~V. 
The LC tank of the VCO is implemented using a custom-designed on-chip spiral inductor, highlighted in the green annotated box shown in Fig~\ref{fig:pll_blocks}. 
A tail current source sets the oscillation amplitude and improves supply noise rejection. 
Finally, a buffer is connected to the VCO to obtain a rail-to-rail output waveform, alongside a similar buffer to balance the differential load.

\begin{table*}
\centering
\footnotesize
\renewcommand{\arraystretch}{1.15}
\setlength{\tabcolsep}{4pt}
\caption{Comparison table of PLL architectures operating near 2.4~GHz band, highlighting key performance metrics.}
\label{tab:pll_arch_comparison}
\resizebox{\textwidth}{!}{
\begin{tabular}{p{4.5cm}p{2.2cm}p{1.5cm}p{2.5cm}p{3.3cm}p{1.8cm}p{1.5cm}}
\hline
\hline
\textbf{PLL Architecture} &
\textbf{VCO} &
\textbf{Process} &
\textbf{Frequency (GHz)} &
\textbf{Phase Noise (dBc/Hz)} &
\textbf{Power (mW)} &
\textbf{Area (mm$^2$)} \\
\hline
\hline

Integer-N PLL \cite{11383160} &
Ring-Oscillator &
\textbf{130\,nm} &
\textbf{2.4} &
-167.88\, (@1MHz) &
11.63&
0.0495 \\
\hline

Integer-N PLL \cite{electronics11071118} &
\textbf{CMOS LC VCO} &
180\,nm &
\textbf{2.4} &
-119\,(@1MHz) &
8&
0.96 \\
\hline


\textbf{CMOS LC-PLL} \cite{11171270} &
\textbf{CMOS LC VCO} &
65\,nm &
10.3 &
-95.12\, (@1MHz) &
6.8 &
---\\
\hline

\textbf{Fractional-N PLL} \cite{11373146} &
 Multi-Core VCO &
\textbf{130\,nm} &
\textbf{0.125--8.4} &
-152.9\, (@10MHz)&
--- &
--- \\
\hline
\textbf{Fractional-N} Oversampling PLL \cite{9537160} &
\textbf{CMOS LC VCO} &
65\,nm &
\textbf{2.4} &
-217.8\, (FOM)&
4.97 &
0.58\\
\hline

\hline

\textbf{Our Design [Fractional-N PLL]} &
\textbf{CMOS LC VCO} &
\textbf{130\,nm} &
\textbf{2.4} &
\textbf{-100.8\,(@1MHz)} &
\textbf{12.73} &
\textbf{0.619} \\
\hline
\hline

\end{tabular}}
\vspace{-0.4cm}
\end{table*}

\subsection{Programmable Fractional-N Divider}

\subsubsection{Dual Modulus Divider (DMD)}

The frequency divider architecture is designed to provide two distinct division ratios of 240 and 248, selectable via the $SEL$ pin. The single-ended output ($OUT_P$) of the VCO is connected to this block as input. The system comprises four stages and one controller block.
\begin{enumerate}
    \item Fixed Divider - The input frequency is initially divided by 4 to ensure a stable square-wave output.
    \item Dual-Modulus Prescaler - Based on the logic state of the controller, the prescaler utilizes pulse swallowing to switch between division ratios of $N=4$ and $N+1=5$.
    \item Mod-7 Synchronous Counter - This stage provides logic signals to the controller block. When cascaded, Stages 2 and 3 provide a total division ratio of either 30 or 31, as determined by the $SEL$ logic.
    \item Duty Cycle Corrector - To ensure signal integrity, the output is processed to achieve a precise 50\% duty cycle, independent of the input pulse width.
\end{enumerate}
The controller block is used to select the divider ratio of the dual-modulus prescaler. 
The inputs to the controller are obtained from the Stage 3 outputs and the $SEL$ pin.

\subsubsection{Shift Register (SR)}

A shift register is employed in the proposed design to reduce the number of I/O pins required by the input interface of the DSM. It is a standard 9-bit shift register with a gated clock and asynchronous active high reset.

\subsubsection{Delta Sigma ($\Delta \Sigma$) Modulator (DSM)}

The main purpose of using a DSM is to push spectral noise, which is generated due to switching between division ratios, to higher frequencies. DSM is designed in such a way that the majority of spectral noise will be pushed beyond the loop bandwidth of the PLL's feedback loop. To this end, we have selected a first-order digital DSM, which consists of a 9-bit signed accumulator, a 9-bit comparator that functions as a 1-bit quantizer, and a mapper that maps the 1-bit output of the quantizer to 9 bits before subtracting from the input.

\begin{figure}[b!]
    \centering
    \vspace{-0.61cm}

    \begin{overpic}[width=1\linewidth]{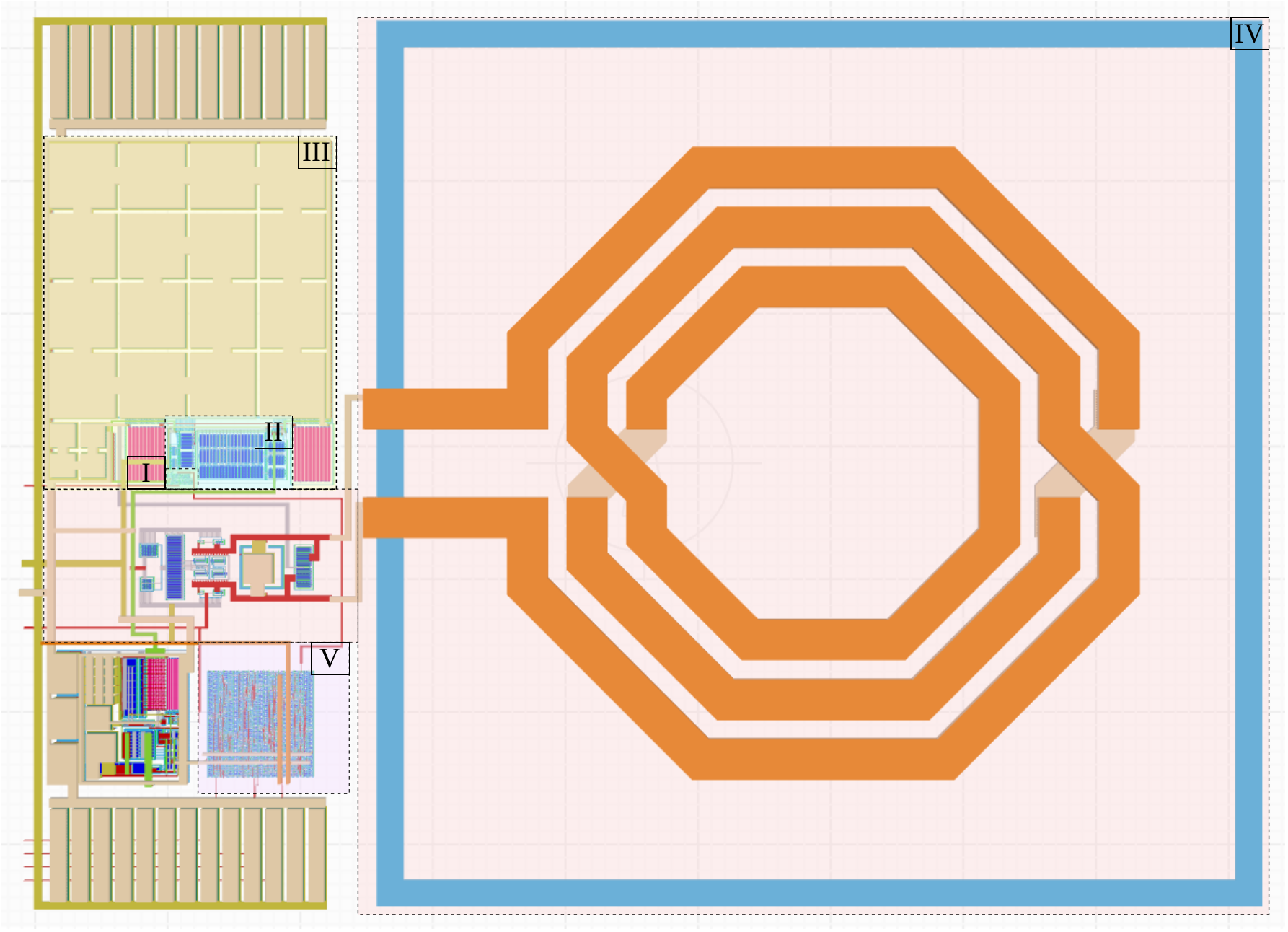}
        \put(-1,0.5){\small\textbf{(a)}}
    \end{overpic}

    \begin{minipage}{0.5\linewidth}
        \centering
        \begin{overpic}[width=\linewidth]{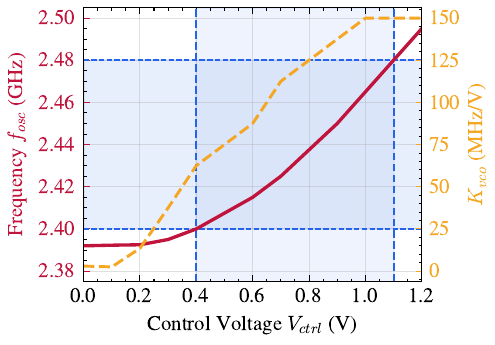}
            \put(-1,0.5){\small\textbf{(b)}}
        \end{overpic}
    \end{minipage}\hfill
    \begin{minipage}{0.5\linewidth}
        \centering
        \begin{overpic}[width=\linewidth]{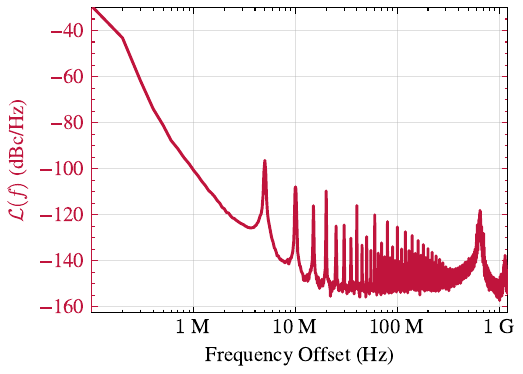}
            \put(0,1.5){\small\textbf{(c)}}
        \end{overpic}
    \end{minipage}

    \caption{ \textbf{Post-layout simulation results of the LC-VCO-based PLL}. (a) Complete layout of the PLL ($930~\mu\text{m} \times 666~\mu\text{m}$). (b) LC-VCO tuning curve demonstrating the variation in sensitivity ($K_{\text{VCO}}$). (c) Phase noise characteristics of the integrated PLL at a carrier frequency of 2.44~GHz.}
    \label{fig:layout}
\end{figure}

\section{Post-Layout Simulation Results}

The proposed LC-VCO-based integrated PLL layout passed both Design Rule Check (DRC) and Layout vs. Schematic (LVS) verification and was validated through post-layout simulations that incorporated parasitic extraction. The complete layout, illustrated in Fig.~\ref{fig:layout}a, occupies a total area of $930~\mu\text{m} \times 666~\mu\text{m}$ ($\approx 0.619~\text{mm}^2$). The LC-VCO tuning range, shown in Fig.~\ref{fig:layout}b, spans 2.4--2.48~GHz, effectively covering the targeted ISM band. The tuning curve's characteristic nonlinearity is attributed to varicap behaviour and layout-induced parasitics, resulting in a sensitivity ($K_{\text{VCO}}$) of approximately 120~MHz/V. The PLL phase noise (Fig.~\ref{fig:layout}c) simulated as outlined in \cite{11383160}, is approximately $-100.8$~dBc/Hz at 1~MHz offset. The reference spur characteristics are around $-40.2$~dBc, which is slightly high, although it still meets BLE specifications, hence further improvements in this aspect are needed. Based on these post-layout simulations, the implemented PLL design successfully maintains stable closed-loop operation and synthesizes the relevant frequencies on channels in the 2.4 GHz ISM band.

\section{Conclusions and Future Works}

This work demonstrates the successful implementation of a fully open-source 2.4~GHz type-II fractional-N PLL in the IHP SG13G2 130~nm BiCMOS process. The main contribution of this work is the design of a cross-coupled differential LC-VCO using an integrated on-chip spiral inductor, which achieves a quality factor ($Q_{\text{diff}}$) of 16.8 at the operating frequency. A performance comparison between the proposed architecture and state-of-the-art PLLs operating in the 2.4~GHz band is summarized in Table~\ref{tab:pll_arch_comparison}. As demonstrated, our open-source design is comparable in performance to existing work utilizing different architectures and to best of our knowledge, is the first demonstration of a fully open-source CMOS PLL design utilizing an integrated on-chip spiral inductor.

Future work will focus on addressing current performance limitations to meet BLE specifications, such as reference spur suppression via charge pump matching and loop filter optimization. Higher-order $\Delta\Sigma$ modulators and fast-locking techniques will also be explored to reduce fractional spurs and settling time while maintaining stable PLL operation.





\bibliographystyle{IEEEtran}
\bibliography{refs}

\end{document}